\documentclass[twocolumn,aps,pra,showpacs]{revtex4}
\usepackage{epsf,epsfig,amsmath,amssymb,graphicx,times}
\begin{document}
\title{Feasibility of continuous-variable quantum key distribution with noisy coherent states}
\author{Vladyslav C. Usenko}
\email{vusenko@bitp.kiev.ua}
\affiliation{Bogolyubov Institute for Theoretical Physics of National Academy of Sciences,
Metrolohichna st. 14-b, 03680, Kiev, Ukraine}
\affiliation{Department of Optics, Palack\' y University, 17. listopadu 50,  772~07 Olomouc, Czech Republic}
\author{Radim Filip}
\email{filip@optics.upol.cz}
\affiliation{Department of Optics, Palack\' y University, 17. listopadu 50,  772~07 Olomouc, Czech Republic}
\date{\today}
\begin{abstract}
We address security of the quantum key distribution scheme based on the noisy modulation of coherent states and investigate how it is robust against noise in the modulation regardless of the particular technical implementation. As the trusted preparation noise is shown to be security breaking even for purely lossy channels, we reveal the essential difference between two types of trusted noise, namely sender-side preparation noise and receiver-side detection noise, the latter being security preserving. We consider the method of sender-side state purification to compensate the preparation noise and show its applicability in the realistic conditions of channel loss, untrusted channel excess noise, and trusted detection noise. We show that purification makes the scheme robust to the preparation noise (i.e., even the arbitrary noisy coherent states can in principle be used for the purpose of quantum key distribution). We also take into account the effect of realistic reconciliation and show that the purification method is still efficient in this case up to a limited value of preparation noise.
\end{abstract}
\maketitle

\section{Introduction}

Secure key distribution \cite{cr} with continuous variables \cite{cm} is a recent practical example of the usefulness of quantum properties exhibited by coherent states generated with ordinary laser systems. The pure coherent states were surprisingly found to be the sufficient resource to distribute a secret key, even through a strongly attenuating channel \cite{RR,exp,coll,Gauss}, while the inefficiency and electronic noise of a homodyne detector in a trusted receiver station does not break security of the key distribution \cite{Lod}. Recently, security analysis of excess noise in the channel and detector was discussed in \cite{Lod2}; moreover, the noise at the remote side was shown to be useful for the scheme security \cite{Cerf}. However, it was also recently shown that excess noise in trusted state preparation can break the security even for a purely lossy channel and ideal homodyne detector \cite{Filip08}. To eliminate this security break, an optimized purification of the prepared noisy coherent states in the trusted sender station was proposed. Interestingly, it was shown that secure communication through any purely lossy channel is possible for arbitrary noisy coherent states, assuming ideal homodyne detection \cite{Filip08}. From this follows that even a noisy coherent state can be a useful quantum resource for secure key distribution in the perfect experimental conditions of pure channel loss. Still an open question is how generally valid is this result for an imperfect homodyne detector, with remaining excess noise in the channel and realistic key reconciliation.

In this article, we discuss security of the continuous-variable (CV) quantum key distribution (QKD) with noisy coherent states through a lossy and noisy channel. The preparation noise is due to the imperfect modulation, which was shown to be destructive for the security of the scheme. We reveal the fundamental difference between various types of trusted noise and show the possibility to establish the secure key transmission with noisy coherent states upon a high level of preparation noise using the local state manipulations in the presence of either trusted detection noise or untrusted channel noise. We also investigate the limits of high modulation variance and show that in this case we can practically eliminate the negative effect of the preparation noise using optimal purification. In addition, we show that the positive purification effect is present also upon limited reconciliation efficiency.

The article is structured as follows: in Sec. \ref{protocol} we briefly recall the QKD with noisy coherent states and mention the effect of the purely lossy channel; in Sec. \ref{noisedet} we add detection noise on the trusted remote party side; in Sec. \ref{channoise} we consider untrusted channel noise; in Sec. \ref{recon} we take into account the effect of the realistic error correction with limited efficiency; and then we finish the article with some concluding remarks.

%%%%%%%%%%%%%%%%%%%%%%%%%%%%%%%%%%%%%%%%%
\section{Coherent states protocol}
\label{protocol}

\begin{figure}
\centerline{\psfig{width=8.0cm,angle=0,file=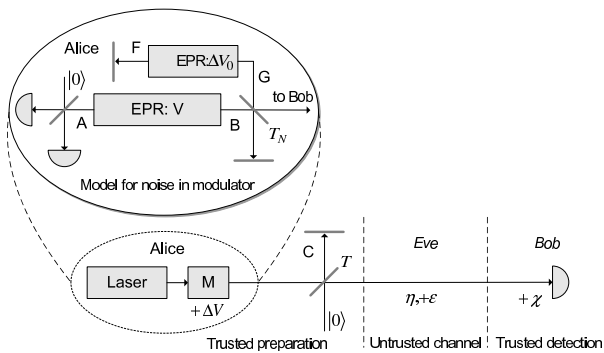}}
\caption{Continuous-variables quantum key distribution scheme based on the noisy coherent states: Laser beam is modulated to encode information in modulator M producing coherent states with superposed additive phase-insensitive noise $\Delta V$. It passes through a variable attenuator with transmittance $T$ (with a second port in vacuum mode $C$) followed by the channel described by excess noise $\epsilon$ and channel transmittivity $\eta$ toward Bob's homodyne detector with additive noise $\chi$. Inlay: equivalent model of Alice noisy state preparation based on the entangled signal source (modes $A,B$) of variance $V$ and entangled noise source (modes $F,G$) of variance $\Delta V$, coupled to signal mode, Alice performing heterodyne measurement on mode B; rest of the scheme is the same as for coherent state modulation.}
\label{scheme}
\end{figure}

CV QKD with coherent states is based on the Gaussian modulation of the states produced by a laser in such a way
that their mixture constitutes a thermal state centered in an origin of phase space and characterized by a source variance $V=1+\sigma$, where $\sigma$ is the modulation variance. Such a setup is referred to as the prepare and measure ($P \& M$) as one of the trusted parties (Alice) for each next
key bit applies modulation to the coherent states, generating them centered at some two Gaussian-distributed values of
quadratures (Fig. \ref{scheme}). The prepared state travels though a quantum channel to the remote trusted party (Bob),
which performs the homodyne measurement of a randomly selected quadrature and stores the result.
Lately, as the sufficient number of preparation and measurement events is carried out, the parties perform key sifting when Bob, using a classical channel,
reveals which of two quadratures he was measuring for each bit and Alice keeps the corresponding bit value, such a scheme
being named reverse reconciliation \cite{RR}. From the obtained classically correlated data, if they they are correlated enough
(which is checked by comparison of the randomly selected subsets),
trusted parties can distill the secure cryptographic key. Physically, the excess noise arises mainly in the
imperfect optical modulator so that each coherent state has the variance $1+\Delta V$, where $\Delta V$ is the preparation noise.
Besides preparation noise, another main source of imperfection is the quantum channel, which has nonunit transmittivity
$\eta$, thus suppressing the signal with losses and adding excess noise $\eta$ to the signal modulation. Also imperfect detection
at Bob's side is assumed, as his homodyne detector has some nonunit efficiency and the excess electronic noise;
we consider imperfect detection as adding excess detection noise $\chi$. Here, for the trusted detector we treat loss and noise in the detection jointly, as the output of the detector is typically classically amplified; thus the detector can be purely noisy so that $\chi$ is the total additive noise of the detector.
At that, the state preparation is assumed to be completely secure (i.e., no attack of an eavesdropper in the sender station is allowed and no side signal leaves the station). Also, the receiver station is trusted, which means that no information is leaking to a potential eavesdropper, whereas the quantum channel is untrusted as it is out of control of the trusted parties. 

It already is well known that untrusted channel noise is limiting the security of CV QKD, whereas trusted detection noise is not contributing to the the knowledge of Eve and is just quantitatively reducing the key rate \cite{RR}. At the same time, it was recently shown that the preparation noise, although being also trusted, is breaking the security of the CV QKD \cite{Filip08}. Thus, in this article we stress out the essential difference between two types of trusted noise, namely preparation and detection, in their effect on the CV QKD security and study the method to compensate the negative effect of trusted preparation noise by applying state purification.
In our theoretical analysis, we model the CV QKD coherent states-based scheme in order to estimate its security, taking realistic values of the parameters of the scheme. In particular, we suppose that the quantum channel has low transmittance $\eta\in (0.01,0.1)$, which corresponds to long-distance optical fibers; the source variance $V$ is mostly taken within the range $V\in (10,100)$, which well corresponds to recent experiments in CV QKD, where the source variance reached approximately $20$ \cite{Lod2} or $40$ \cite{RR} shot-noise units. The typical values of the homodyne detector electonic noise in the mentioned experiments were $0.041$ and $0.33$, respectively, but the limited detection efficiency makes the overall homodyne detection noise higher. 

At the same time, it is hard to assess the possible values of the preparation noise, as the sources of this noise may be essentially different; it can be either imperfect modulation of a shot-noise limited coherent source or the increased variance of a laser, which is used to produce the exact state, displaced according to a next bit value. Besides, the values of such noise were not discussed in experimental CV QKD as this noise is difficult to estimate and calibrate \cite{Lod2}. Hence all the noise in the receiving station was considered to be the impact of Eve and the sources were usually assumed as ideal, focusing on other possible imperfections. However, the simplest sources for a classical coherent optical communication link are cheap laser diodes together with the usual low-frequency integrated amplitude and phase modulators \cite{Betti}. While such a source still has enough temporal coherence to build the coherent communication link, its intensity and phase noise added at the lower modulation frequencies makes the source to be not shot-noise limited, exhibiting additional excess noise \cite{Bachor}. For the classical coherent communication, this is not a serious problem, and these issues can be considered to be just technical, but it appears to be essential for QKD, as the preparation noise can quickly lead to a security break. Hence, we propose to distinguish between trusted preparation noise and the rest of the noise and show that this may improve the applicability of CV QKD as we can compensate security-breaking preparation noise. Note, that in the case of the discrete-variable coding the mixed signal states were considered already in \cite{Koashi}.

It is worth mentioning, that the method of sender-side attenuation, which we propose and investigate in this article in order to purify the states and make CV QKD more robust to the preparation noise, is most likely just one of the possible ways to reduce the negative effect of the state preparation noise. Yet we propose it as the simplest method to purify the states by using just a linear optical device, the beam-splitter, in case the noisy source is given and cannot be otherwise improved. In addition, by introducing attenuation $T$, we can clearly show the effect analytically and optimize it as well as estimate the limitations, imposed on the purification method by the reconciliation efficiency.

As a matter of fact, the attenuation of the signal on the trusted source side was already applied in \cite{Lod2} without the analysis of the impact. Its effect is actually twofold: to purify the noisy states by suppressing the noise and to optimize the signal-to-noise ratio for the given reconciliation efficiency. Still, the purifying effect was not theoretically analyzed. Moreover, the attenuation of laser diodes was proposed by decreasing their power \cite{Lod}, but this method is not applicable for the case, when modulation is performed directly in the laser, not by a standalone modulator. Also, the feedback systems were applied \cite{RR,Lod2} to control the level of modulation and they can be used to reduce the noise, but as their effect is nonlinear, it is not considered in this article. 

%%%%%%%%%%%%%%%%%%%%%%%%%%%%%%%%%%%%%%%%%%%%%%%%%%%%%%%%%%%%%%%%%%%%%%%%%%%%%%%%%%%%%%%%%%%
%%%%%%%%%%%%%%%%%%%%%%%%%%%%%%%%%%%%%%%%%%%%%%%%%%%%%%%%%%%%%%%%%%%%%%%%%%%%%%%%%%%%%%%%%%%%

For the security analysis, it is assumed that Eve couples signal pulses to her ancillary states and stores them in a quantum memory. Then she can perform either an individual attack after the key sifting procedure, when she measures her ancillary states individually, or a collective attack during the key distillation, when she applies the optimal collective measurement on the stored ancillary states. We limit our study of security of the scheme by Gaussian collective attacks as they were shown to be optimal \cite{Gauss} and keep to the the recent proof, that, similar to the case of discrete-variable QKD, the protocol is secure against the most general attacks, if it is secure against the collective ones \cite{Renner}. For individual attacks, in case of reverse reconciliation, the security analysis is based on the fact that maximum information available to Eve is bounded by the classical (Shannon) mutual information, characterizing her knowledge of Bob's data, while in the case of collective attacks Eve is possessing the quantum (von Neumann) information on Bob's measurement results, which is limited by the Holevo bound; in all cases, the criterion for security is formulated as the exceeding of mutual information between Alice and Bob over the information available to Eve on Bob's data; under this condition the secure key can be distilled by the standard procedures. Thus, for the collective attack scenario and one-way
classical postprocessing, there is always a key distillation
protocol generating secure key with minimal rate
\begin{equation}\label{gen}
I=I_{AB}-\chi_{BE},
\end{equation}
where $I_{AB}$ is Shannon mutual information from data obtained by Alice and Bob,
$\chi_{BE}$ is the Holevo information quantity between receiver $B$ and
potential eavesdropper Eve \cite{DV}. The Holevo quantity can be
written as $\chi_{BE}=S_E-\int P(B)S_{E|B}dB$, where $S_E$ is von Neumann
entropy of the eavesdropper's state $\rho_E$ \cite{Gauss}. The quantity
$S_{E|B}$ is the von Neumann entropy of the eavesdropper state
$\rho_{E|B}$ conditioned by the receiver measurement result $B$, and $P(B)$ is the distribution of the measured results.
If $I>0$, then information shared by Alice and Bob is larger than information accessible to an eavesdropper which is performing collective attacks. Then optimal distillation procedure can generate a secure key between trusted parties. 

A weaker condition is for security against only the individual attacks. In this case,
the lower bound on secure key rate is given $I_i=I_{AB}-I_{BE}$,
where $I_{BE}$ is Shannon mutual information between the
eavesdropper and the receiver, if the eavesdropper applies an optimal strategy.
Remarkably, for ideal coherent-state preparation without any excess
noise, it is always possible to achieve $I^{max}=-\log_2(1-\eta)/2>0$ for an arbitrary lossy channel
(but without excess noise)  with $\eta>0$ and for arbitrary modulation
variance $\sigma > 0$ \cite{coll}. 

While we consider the security of the scheme for collective attacks, we will use the individual attack case to estimate the regions of parameters, where the scheme already becomes insecure for any attacks, because insecurity against individual attacks automatically means insecurity against the collective, as the latter are more effective. Furthermore, the individual attack case enables us to analytically show the effect of state purification as well as constitute the limits to which collective attacks tend in the high modulation regime.

So, we investigate the security of the scheme based on the noisy coherent states in different realistic conditions. The situation, when only the state preparation noise is present in the CV QKD based on the coherent states and the quantum channel is purely lossy, was already addressed in \cite{Filip08}. It was shown that the preparation noise is strongly destructive for security of the scheme, but the purification on the side of the trusted source can provide the possibility to establish secure key distribution upon any value of the preparation noise, and that security is guaranteed against both individual and collective attacks. Although seeming contradictory at first glance, this result can be clearly explained from the physical considerations. As the state coherence is preserved upon any attenuation, by suppressing the noisy coherent state we suppress the noise, but the state coherence remains, so that while the intensity becomes lower and the key rate decreases, the state becomes pure enough to provide the security of the key distribution, which without purification could not be possible. The earlier work provided the proof of principle for suppressing preparation noise with purification, while in this article we perform the feasibility check of the method in the realistic conditions of trusted detection noise, untrusted channel noise, and imperfect key reconciliation. 

Generally we investigate whether a noisy source can be used for a CV coherent states-based QKD given that the source cannot be replaced with a better one. While the exact level of preparation noise is the question of cost and complexity of the state preparation equipment, in this article we rather generally discuss the impact of the preparation noise on the QKD security, than refer to a specific experimental arrangement. In this sense, our work is a step from perfect laboratory conditions toward the realistic CV QKD, based on the affordable signal sources.

\section{Noiseless channel and trusted detection noise}
\label{noisedet}

First we generalize the result presented in \cite{Filip08} that is valid for the lossy channel and ideal homodyne detector. Keeping the noiseless channel in mind, we assume a realistic lossy and noisy trusted homodyne detector. This represents an optimistic scenario of realistic experimental setup, taking into account the minimal impact of natural channel background noise into narrow-band homodyne detection.

First we calculate the impact of noise on the security against individual attacks to estimate the insecurity region. As we perform the calculations in the equivalent entangled-based source setup \cite{Lod2}, the overall expression for the lower bound on the key rate, which is secure against individual attacks using the reverse reconciliation, is \cite{RR}
\begin{equation}
\label{indkeyrgen}
I_i=\frac{1}{2}\log_2{\frac{V_{B|E}}{V_{B|A^M}}},
\end{equation}
where $V_{B|E}=V_B-\frac{C_{BE}^2}{V_E}$ and $V_{B|A^M}=V_B-\frac{C_{AB}^2}{V_A+1}$ are the relevant conditional variances. The one unit of noise added to $V_A$ appears from the heterodyne measurement on Alice's side. In our case, the variances are $V_A=V$, $V_B=\eta(V+\Delta V)+1-\eta+\chi$ and $V_E=(1-\eta)(V+\Delta V)+\eta$ and the mode correlations are $C_{BE}=\sqrt{\eta(1-\eta)}(1-V-\Delta V)$ whereas $C_{AB}=\sqrt{\eta (V^2-1)}$. Detection noise $\chi$, being out of control by the eavesdropper, is not involved in the correlation $C_{BE}$; hence, it does not break the security but only lowers the key rate. The conditional variances are $V_{B|A^M}=1+\eta\Delta V+\chi$ and
\begin{equation}
V_{B|E}=\frac{1}{\frac{\eta}{V+\Delta V}+1-\eta}+\chi;
\end{equation}
then from the explicit expression (\ref{indkeyrgen}) for the lower bound on key rate in this case we can derive the security constraint on the level of the preparation noise:

\begin{equation}
\label{maxdvind}
\Delta V < \frac{1}{2}-\frac{V}{2}+\sqrt{(V-1)\bigg(V-1+\frac{4}{1-\eta}\bigg)},
\end{equation}

which does not depend on the detection noise, added on the remote receiving side.

In the limit of arbitrary large source variance (i.e., arbitrary high modulation), this constraint turns to 

\begin{equation}
\label{maxdvindlim}
\Delta V < \frac{1}{1-\eta}, 
\end{equation}

which means that for strongly attenuating channels $\eta << 1$, even if the modulation is arbitrarily high, the preparation noise should not exceed one shot noise unit, while this condition becomes more restrictive for the realistic cases of finite modulation. 

In the case of collective attacks, as mentioned, Eve's information on the key is limited by the Holevo quantity.
Since we are considering a purely attenuating channel, we substitute it by a beam splitter with the transmittance equivalent to the channel transmittivity $\eta$. We perform the straightforward calculations of the Holevo quantity $\chi_{BE}=S_E-\int P(B)S_{E|B}dB$, bounding Eve's quantum information from a state going from the beam splitter to Eve. Since we are working with Gaussian states, entropy $S_{E|B}$ does not depend on the Bob's measurement result and we can simply use $\chi_{BE}=S_E-S_{E|B}$. Using the expression for von Neumann entropies \cite{entropy}, the Holevo quantity is calculated from
\begin{equation}\label{holevo1}
\chi_{BE}=G(\frac{\lambda_1-1}{2})-G(\frac{\lambda_2-1}{2}),
\end{equation}
where $G(x)=(x+1)\log (x+1)-x\log x$, $\lambda_1$ is the symplectic eigenvalue of Eve's mode covariance matrix $\gamma_E=V_E\mathbb{I}$, in fact $\lambda_1=\sqrt{\mbox{Det}\gamma_E}$ and $\lambda_2$ is the symplectic eigenvalue of the covariance matrix $\gamma_E^{x_B}$ characterizing the state of Eve's mode after Bob's projective measurement:
\begin{equation}
\gamma_E^{x_B}=\gamma_E-\sigma_{BE}(X \gamma_B X)^{MP}\sigma_{BE}^T,
\end{equation}
where $\gamma_B=V_B\mathbb{I}$ is the covariance matrix of Bob's mode, $\sigma_{BE}$ characterizes correlation between Bob's and Eve's modes, MP stands for Moore Penrose inverse of a matrix (also known as pseudoinverse) and
\begin{equation}
X =
\left( \begin{array}{cc}
1 & 0 \\
0 & 0
\end{array} \right).
\end{equation}
Finally, the mutual information $I_{AB}$ is calculated using Shannon entropies as
\begin{equation}
I_{AB}=\frac{1}{2}\log_2{\frac{V_{A}+1}{V_{A|B}+1}}.
\end{equation}

\begin{figure}
\centerline{\psfig{width=8.0cm,angle=0,file=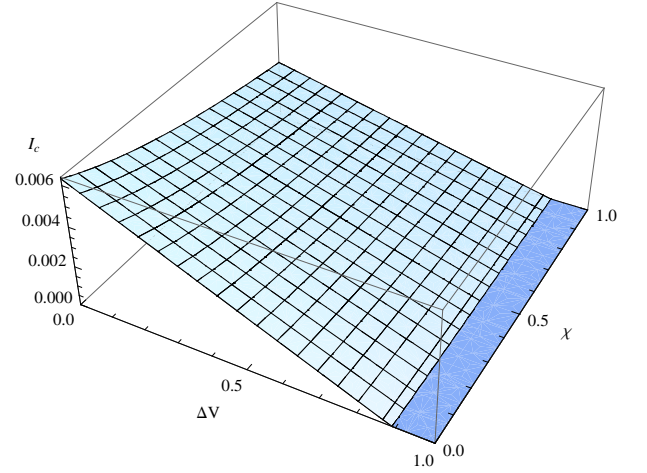}}
\caption{The effect of trusted detection noise $\chi$ and trusted preparation noise $\Delta V$ on the secret key in the case of no purification. Security against collective attacks is considered, source variance $V=20$, and channel transmittivity $\eta=0.01$.}
\label{trusted_noise}
\end{figure}

The explicit expression for the key rate is obtainable analytically, but it is too lengthly. However, similarly to the case of individual attacks, the security of the scheme is limited by the preparation noise and the boundary for collective attacks is close to that of the individual attacks (\ref{maxdvind}), while in the limit of arbitrary high modulation it exactly coincides with (\ref{maxdvindlim}). The effect of the preparation noise in comparison to the effect of the trusted detection noise in case of collective attacks is shown in Fig. \ref{trusted_noise} for the lossy channel with $\eta=0.01$ and realistic source variance $V=20$. One may see, that despite the fact, that both kinds of noise are trusted, the preparation noise is security breaking, unlike the detection noise, which is only quantitatively reducing the key rate -- the effect, which we observed above for the individual attacks case.

%%%%%%%%%%%%%%%%%%%%%%%%%%%%%%%%%%%%%%%%%%%%%%%%%%%%%%%%%%%%%%%%%%%%%%%%%%%%%%%%%%%%%%%%%%%
%%%%%%%%%%%%%%%%%%%%%%%%%%%%%%%%%%%%%%%%%%%%%%%%%%%%%%%%%%%%%%%%%%%%%%%%%%%%%%%%%%%%%%%%%%%%

Now let us assume that purifying attenuation is applied to the signal states prior to sending them through the quantum channel. By supposing the attenuator transmittivity is $T$, we can see its positive affect already for individual attacks, as the lower bound on the key rate in this case is 

\begin{eqnarray}
\label{indkr}
I_i=\frac{1}{2}\log_2{\bigg(
\frac{1+T(V+\Delta V -1)}{1+T(V+\Delta V -1)(1-\eta)}+\chi\bigg)}-{}
\nonumber \\
{}-\frac{1}{2}\log_2{(1+T\eta\Delta V+\chi)} &
\end{eqnarray}

and from this expression it is clear that attenuation $T$ is suppressing the preparation noise $\Delta V$. If the modulation variance is arbitrary high, this effect is even more evident, as the key rate turns to

\begin{eqnarray}
\label{indkrlim}
I_i\big|_{V \to \infty}=\frac{1}{2}\log_2{\bigg(\frac{1}{1-\eta}+\chi\bigg)}-{}
\nonumber \\
{}-\frac{1}{2}\log_2{(1+T\eta\Delta V+\chi)}, &
\end{eqnarray}
while the preparation noise $\Delta V$ threshold becomes
\begin{equation}
\label{inddvmax}
\Delta V_{i,max}\big|_{V \to \infty}=\frac{1}{T(1-\eta)}
\end{equation}
and can be made arbitrary high by purification.

For the finite $V$, if we take the derivative of the key rate (\ref{indkr}) by purifying attenuation $T$ in the point $T=0$, we obtain
\begin{equation}
\label{indder}
\frac{dI_{i}}{dT}\bigg\arrowvert_{T=0}=\frac{1}{\log{4}}\frac{\eta(V-1)}{1+\chi},
\end{equation}
which is always positive as $V \ge 1$. This means that close to $T=0$ there is always some $T>0$ that is sufficient to provide with the secure key rate because at point $T=0$, the key rate is equal to 0.
From the given expressions it is evident that detection noise, being not under the control of an eavesdropper, only decreases the key rate but does not destroy the transmission, even for finite source variance $V$, the detection noise $\chi$ does not change the security bounds, whereas preparation noise $\Delta V$ plays crucial role.
It is also worth mentioning that while secure key transmission is possible for arbitrarily high preparation noise upon arbitrary strong purification, for the given parameters, there always exists an optimal purification level, which maximizes the secure key rate. We skip the corresponding lengthly equations here, but in the further exposition we calculate the maximal key rate values in various conditions.

If the optimal purification is applied on the noisy signal and the modulation is extremely intense so that the source variance $V \to \infty$, then the expression for the bound on the key rate for individual attacks (\ref{indkr}) analytically turns to the expression for the case when the preparation is pure, i.e. $\Delta V=0$ and the modulation is infinitely high:

\begin{equation}
I_i\big|_{V \to \infty}^{T=T_{opt}}=
\frac{1}{2}\log{\frac{1+\chi(1-\eta)}{(1+\chi)(1-\eta)}}=I_i\big|_{V \to \infty}^{\Delta V=0},
\end{equation}

which means that by combining optimal purification with large modulation variance we can completely compensate the negative effect of preparation noise and achieve the values of key rate corresponding to the generation of pure states. Furthermore we will obtain a similar result in other cases.

\begin{figure}
\centerline{\psfig{width=8.0cm,angle=0,file=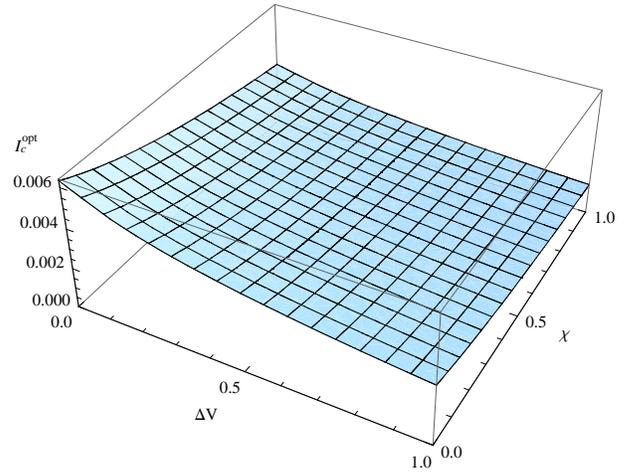}}
\caption{The effect of trusted detection noise $\chi$ and trusted preparation noise $\Delta V$ on the secret key if the optimal purification is applied. Security against collective attacks is considered, source variance $V=20$, and channel transmittivity $\eta=0.01$.}
\label{trusted_noise_opt}
\end{figure}

In the case of collective attacks, similarly to the individual attacks, we can calculate the derivative of the key rate by purifying attenuation $T$ and evaluate it in the limit of $T\to 0$, arriving at the same result as (\ref{indder}) for individual attacks, proving that for finite modulation variance $\sigma$, we can always find some $T$ close to 0, which will provide secure transmission for any preparation noise $\Delta V$ as $I_{c}\arrowvert_{T\to 0}=0$. This result was also confirmed  numerically by estimation of the maximal key rate secure against collective attacks upon optimal attenuator setting for given preparation noise and channel loss level. The optimized key rate versus preparation and detection noise is presented in Fig. \ref{trusted_noise_opt}. It is evident from the plot that by applying optimal purification to the noisy states we can turn the negative security-breaking effect of the preparation noise to the same as that of the detection noise, which is only reducing the key rate, but not turning it to zero.

%%%%%%%%%%%%%%%%%%%%%%%%%%%%%%%%%%%%%%%%%%%%%%%%%%%%%%%%%%%%%%%%%%%%%%%%%%%%%%%%%%%%%%%%%%%
%%%%%%%%%%%%%%%%%%%%%%%%%%%%%%%%%%%%%%%%%%%%%%%%%%%%%%%%%%%%%%%%%%%%%%%%%%%%%%%%%%%%%%%%%%%%

In the limiting case of infinitely high modulation (i.e., source variance $V \to \infty$), the expressions for key rates and preparation noise thresholds are the same as for individual attacks, namely (\ref{indkrlim},\ref{inddvmax}), which means that by combining optimal attenuation and arbitrary high modulation, we can completely remove the negative effect of the preparation noise.

\section{Channel noise and untrusted detection noise}
\label{channoise}

While state purification was shown to be effective in case when trusted noise is present in the system, let us estimate the applicability of the method in the worst-case scenario, considering that noise, observable in the receiving station, is not trusted. This can be channel background noise in the case of broad-band detectors, untrusted detection noise or the influence of Eve, but as channel is the only untrusted part of our scheme, we refer to this noise as to the channel one.

Let us consider excess noise $\epsilon$, which is out of control of trusted parties. In the case of individual attacks, we use the "entangling cloner" scheme \cite{RR}, which allows Eve to perform optimal eavesdropping by substituting the channel of loss $\eta$ and excess noise $\epsilon$ with her apparatus. The optimality of the cloner attack is due to the fact that it enables Eve to reach the limit of her knowledge on Bob's data, described by the Heisenberg uncertainty principle, with the statement being valid under the assumptions that the used states are Gaussian and Eve knows the parameters of the channel, which well holds in our case. The attack uses pure two-mode squeezed vacuum state having local variance $N$; one mode is reflected to Bob's mode through the beam splitter with reflectivity $1-\eta$, whereas the transmitted part is kept by Eve. The second mode from the two-mode squeezed vacuum state is purely in the hands of Eve. Eve is fixing $N$ in order to satisfy $N=1+\frac{\eta \epsilon}{1-\eta}$, this way emulating the channel by her apparatus. She can store both of her modes in quantum memories and measure the appropriate quadratures of both modes in order to decrease the noise added to the measured part of the signal mode upon eavesdropping.

If Eve's modes variances before interaction with the signal are $V_{E^0_1}=V_{E^0_2}=N$ and Bob's mode is $V_{B_0}=T(V+\Delta V)+1-T$, then after interaction between modes $B$ and $E_1$ their variances are $V_B=\eta V_{B_0}+1-\eta+\eta\epsilon$ and $V_{E_1}=\eta V_{E^0_1}+(1-\eta)V_{B_0}$, while correlation between the modes is $C_{BE_1}=\sqrt{\eta(1-\eta)}(V_{E^0_1}-V_{B_0})$.
By calculating conditional variance as $V_{B|E_1}=V_B-C^2_{BE_1}/V_{E_1}$ and taking into account that after Eve's measurement on mode ${E_2}$ the conditional variance of measurement on mode ${E_1}$ becomes $V_{E_1|E_2}=1/N$, we obtain the expression for conditional variance of Eve's results on Bob's measurement results, which is

\begin{equation}
V_{B|E}=V_{B|E_1E_2}=\frac{1}{\eta\big(\frac{1}{T(V+\Delta V)+1-T}-1+\epsilon\big)+1}
\end{equation}

whereas Alice's uncertainty from the measured results is

\begin{equation}
V_{B|A^M}=1+\eta T \Delta V+\eta\epsilon.
\end{equation}

From the expression (\ref{indkr}) we can explicitly obtain the lower bound on the key rate in case of the untrusted channel noise:

\begin{eqnarray}
\label{indkrcn}
I_i=\frac{1}{2}\log_2{\bigg(
\frac{1}{\eta\big(\frac{1}{T(V+\Delta V)+1-T}-1+\epsilon\big)+1}\bigg)}-{}
\nonumber \\
{}-\frac{1}{2}\log_2{(1+T\eta\Delta V+\epsilon\eta)},
\end{eqnarray}

The derivative of the key rate by $T$ in the point $T=0$ is again always positive with $\frac{1}{\log_{10}{4}}\frac{\eta(V-1)}{1+\eta \epsilon}$, but as the key rate at $T=0$ is equal to $0$ only when $\epsilon=0$, being negative at any other $\epsilon>0$, the security upon channel noise is no longer guaranteed only by purification.

However, we may estimate the region of channel loss in which the purification can provide security for the given noise levels. As the channel excess noise can be hardly detected by trusted parties, in the entanglement cloner scenario, we take Eve's Einstein-Podolsky-Rosen (EPR) source variance $N$ as the parameter, which is independent on $\eta$. Within this approach we may estimate the bound on $\eta$ which restricts the security of key distribution for the given noise levels using the condition for optimal purification $T_{opt}<1$, this bound on security can be expressed as

\begin{equation}
\label{etamax}
\eta < \frac{1}{1+\frac{V-1}{N(1-V-\Delta V+\Delta V(V+\Delta V)^2)}},
\end{equation}

it can be compared to the similar bound on $\eta$, restricting the security of the key upon no purification, $T=1$. We skip the lengthy expression for the latter bound, but the calculations show that the region (\ref{etamax}) is no larger than the similar region for no trusted side attenuation, i.e. purification provides security of the key for any $\Delta V$ in the same region of parameters, where the security was provided for no purification and no preparation noise.

The optimal purification, which maximizes the secure key rate upon given parameters can be expressed as

\begin{eqnarray}
\label{topt}
\lefteqn{T_{opt}=\frac{1}{V+\Delta V-1}\times{}}
\nonumber \\
&&{}\times\Bigg(\sqrt{\frac{(V+\Delta V-1)(\eta\epsilon+1)-\eta\Delta V}{\Delta V(\eta\epsilon+1-\eta)}}-1\Bigg)
\end{eqnarray}

In the limit of infinitely high modulation (i.e., source variance $V \to \infty$), the bound on the key rate turns to

\begin{eqnarray}
\label{indkrlim2}
I_i\big|_{V \to \infty}=\frac{1}{2}\log_2{\frac{1}{\eta\epsilon+1-\eta}}-{} 
\nonumber \\
{}-\frac{1}{2}\log_2{(1+T\eta\Delta V+\eta\epsilon)}, &
\end{eqnarray}

while the threshold on the preparation noise involves channel noise, which is additionally limiting the tolerable preparation noise:

\begin{equation}
\Delta V_{i,max}=\frac{1-\epsilon}{T(1-\eta+\eta\epsilon)}-\frac{\epsilon}{T}.
\end{equation}

However, if the optimal purification (\ref{topt}) is applied in this case, the expression for the secure key rate turns to the one for the case of infinite modulation and no preparation noise:

\begin{eqnarray}
I_i\big|_{V \to \infty}^{T=T_{opt}}=\frac{1}{2}\log_2{\frac{1}{\eta\epsilon+1-\eta}}-{} 
\nonumber \\
{}-\frac{1}{2}\log_2{(1+\eta\epsilon)}=I_i\big|_{V \to \infty}^{\Delta V=0}, &
\end{eqnarray}

which means that the combination of optimal purification and sufficiently large modulation variance can completely eliminate the negative effect of the preparation noise. Still, for the limited source variance, the security can be provided for any given preparation noise within the limited range of noise introduced by an eavesdropper as described by (\ref{etamax}).

In the case of collective attacks performed by an eavesdropper we can no longer use direct calculations of the Holevo quantity used in the previous section. Rather, we assume that Eve can purify the complete state shared among all the parties. In order to perform calculations for this case we must switch to the entangled-based scheme, which was shown to be equivalent to the prepare-and-measure one \cite{Lod2} and take into account all the trusted modes in Alice's station.
Similarly, we will assume the worst-case detection scenario, that all the noise and loss in the Bob's station is untrusted.

In the entanglement-based scheme, the entanglement is generated in the source $EPR:V$ by mixing two pure orthogonally squeezed states with the squeezing variance $V_S$ at a balanced beam splitter producing two entangled modes A and B.  The mode A in the thermal state of variance $V=(V_S+V_S^{-1})/2$ is measured by Alice simultaneously in both quadratures by the heterodyne detector, while mode B is coupled to the noise mode G from the similar entangled source $EPR:\Delta V_0$ of variance $\Delta V_0$, emulating preparation noise in modulator (see Fig.~\ref{scheme}, inlay).
To emulate the additive noise with noise variance $\Delta V$, the coupling between $B$ and $G$ is strongly asymmetrical, with almost unit transmittivity $T_N\approx 1$. This setup corresponds to the preparation noise of variance $\Delta V=(1-T_N)\Delta V_0$.

Now the state of ABCFG is pure in the absence of channel noise and we can use the fact that Eve purifies this state so that von Neumann entropy $S_E=S_{ABCFG}$. Then, after Bob's measurement the system ACFG is pure and the conditional entropy $S_{E|B}=S_{ACFG|B}$. Thus, the Holevo quantity becomes $\chi_{BE}=S_{ABCFG}-S_{ACFG|B}$. We perform calculation of these values similarly to the case of individual attacks, by obtaining symplectic eigenvalues of the five-mode covariance matrix $ABCFG$ and four-mode matrix $ACFG|B$, which can be done purely numerically. As a result, we get the method allowing us to perform calculations of the lower bound of the key rate upon any conditions in the presence of both preparation and channel excess noise and we investigate the effect of purification on the applicability of the scheme. Moreover, we have numerically confirmed that in the absence of channel noise this method, based on the state purification, is equivalent to the method based on Eve's mode calculation, which was used in the previous section.

\begin{figure}
\centerline{\psfig{width=8.0cm,angle=0,file=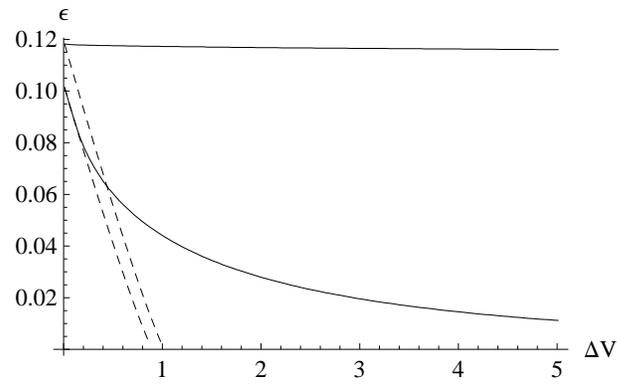}}
\caption{Maximal tolerable channel excess noise versus given preparation noise for no Alice-side purification (dashed line) and with optimal attenuator setting (solid line), source variance $V=10$ (lower lines) and $V=10^5$ (upper lines), and channel loss $\eta=0.01$}
\label{channel_noise}
\end{figure}

Like in the individual attacks scenario, in the case of the collective attacks, security is now limited by both preparation and channel excess noise. We calculated the maximal tolerable channel excess noise as the function of preparation noise both for optimal purification and no purification on Alice's side; the results are given in Fig. \ref{channel_noise} for typical source variance of $V=10$ and for almost ideal case of extremely large source variance $V=10^5$. It was checked numerically and can be seen from the graphs that, similarly to the case of individual attacks, when the purification is optimized, the transmission is now possible upon arbitrarly high preparation noise if the channel itself was not destructive, and the scheme is not becoming more sensitive to the channel noise than without purification. Also for large source variance $V$ the dependence of maximal tolerable excess noise on given state preparation noise approaches saturation, starting to be very slow. 

Thus, we need to increase source variance $V$ (by increasing modulation variance $\sigma$) in order to weaken the dependence of key rate on the preparation noise $\Delta V$. However, for small source variance $V \to 1$, the dependence on preparation noise is qualitatively the same, just the scheme quickly becomes sensitive to channel noise.

%%%%%%%%%%%%%%%%%%%%%%%%%%%%%%%%%%%%%%%%%%%%%%%%%%%%%%%%%%%%%%%%%%%%%%%%%%%%%%%%%%%%%%%%%%%
%%%%%%%%%%%%%%%%%%%%%%%%%%%%%%%%%%%%%%%%%%%%%%%%%%%%%%%%%%%%%%%%%%%%%%%%%%%%%%%%%%%%%%%%%%%%

\begin{figure}
\centerline{\psfig{width=8.0cm,angle=0,file=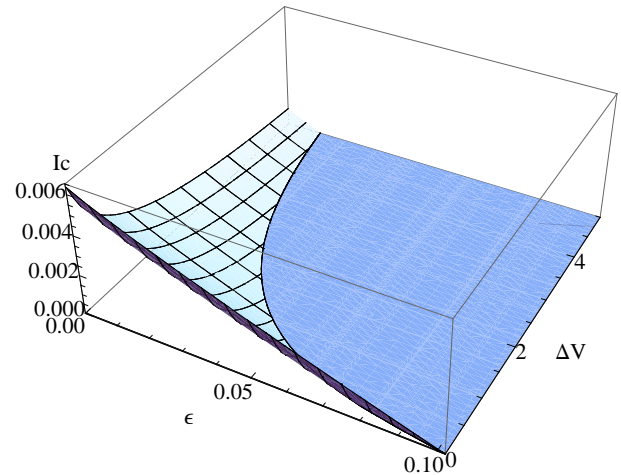}}
\caption{Maximal key rate as the function of channel excess noise $\epsilon$ and preparation noise $\Delta V$, source variance $V=10$, and channel loss $\eta=0.01$}
\label{KR_3D_10}
\end{figure}

\begin{figure}
\centerline{\psfig{width=8.0cm,angle=0,file=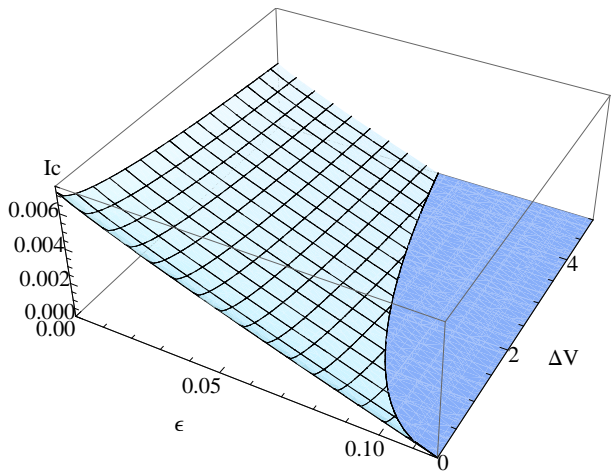}}
\caption{Maximal key rate as the function of channel excess noise $\epsilon$ and preparation noise $\Delta V$, source variance $V=100$, and channel loss $\eta=0.01$}
\label{KR_3D_100}
\end{figure}

For the realistic cases, in order to perform the qualitative comparison, the maximal key rate was calculated as the function of preparation noise and channel excess noise and is given in Fig. \ref{KR_3D_10} for $V=10$ and in Fig. \ref{KR_3D_100} for $V=100$. It is evident from the graph, that for relatively small channel noise the security of the scheme can be provided upon any preparation noise level, while as the channel noise increases, the security is not guaranteed anymore.

%%%%%%%%%%%%%%%%%%%%%%%%%%%%%%%%%%%%%%%%%%%%%%%%%%%%%%%%%%%%%%%%%%%%%%%%%%%%%%%%%%%%%%%%%%%
%%%%%%%%%%%%%%%%%%%%%%%%%%%%%%%%%%%%%%%%%%%%%%%%%%%%%%%%%%%%%%%%%%%%%%%%%%%%%%%%%%%%%%%%%%%%

%%%%%%%%%%%%%%%%%%%%%%%%%%%%%%%%%%%%%%%%%%%%%%%%%%%%%%%%%%%%%%%%%%%%%%%%%%%%%%%%%%%%%%%%%%%
%%%%%%%%%%%%%%%%%%%%%%%%%%%%%%%%%%%%%%%%%%%%%%%%%%%%%%%%%%%%%%%%%%%%%%%%%%%%%%%%%%%%%%%%%%%%

\begin{figure}
\centerline{\psfig{width=8.0cm,angle=0,file=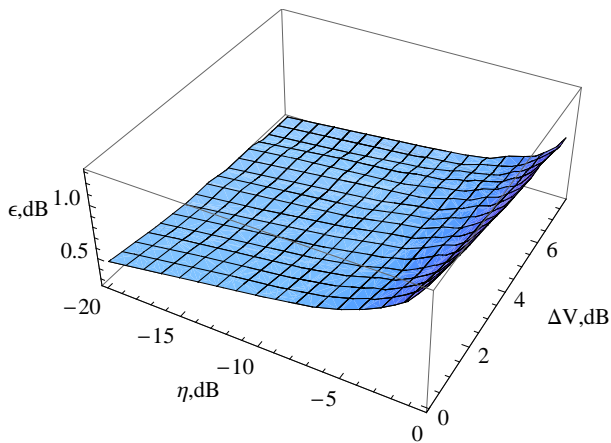}}
\caption{The region of parameters for which the security is guaranteed for source variance $V=100$}
\label{security_surface-dB}
\end{figure}

The region of parameters for which the security is guaranteed is plotted for source variance $V=100$ at Fig. \ref{security_surface-dB} and one can see that over the broad values of channel loss $\eta$ and channel noise $\epsilon$, the security is guaranteed for any preparation noise $\Delta V$, while dependence on $\Delta V$ upon its high values becomes more linear and flat, approaching the horizontal surface for extremely high source varance $V\to \infty$.

%%%%%%%%%%%%%%%%%%%%%%%%%%%%%%%%%%%%%%%%%%%%%%%%%%%%%%%%%%%%%%%%%%%%%%%%%%%%%%%%%%%%%%%%%%%
%%%%%%%%%%%%%%%%%%%%%%%%%%%%%%%%%%%%%%%%%%%%%%%%%%%%%%%%%%%%%%%%%%%%%%%%%%%%%%%%%%%%%%%%%%%%

Let us now consider the relatively simple case of no preparation noise (i.e., $\Delta V=0$). The explicit expressions for the key rate even in this case are lengthy [although analytically obtainable through the symplectic eigenvalues of matrices $AB$ and $A|x_B$ and extended expression (\ref{holevo1})]. We are interested in extremely lossy channels, so we perform a series expansion of the key rate with respect to source variance $V$ around infinity and to channel loss $\eta$ around zero. Such expansion up to the first order has the form

\begin{eqnarray}
I_c\big|^{\Delta V=0}\approx\frac{\eta}{\log{4}}
\Big(1-\epsilon+\epsilon\log{\frac{\eta\epsilon}{2}}\Big)+O[\eta]^{3/2}+{} 
\nonumber \\
{}+\frac{1}{V}\Big(-\frac{2\eta}{\log{8}}+O[\eta]^{3/2}\Big)+O\Big[\frac{1}{V}\Big]^{3/2}, &
\end{eqnarray}

while for $V \to \infty$ the approximate normalized expression for the key rate turns to

\begin{equation}
\label{approxseries}
I_c\big|_{V \to \infty}^{\Delta V=0}\approx 0.721\eta-1.221\eta\epsilon+0.721\eta\epsilon\log{\eta\epsilon}.
\end{equation}

Now we perform the analytical least-squares fit of the numerical data \cite{hudson}, obtained for the optimal purification when preparation noise is present:

\begin{equation}
\label{approxnum}
I_c\big|_{V \to \infty}^{T=T_{opt}}\approx 0.722\eta-1.237\eta\epsilon+0.731\eta\epsilon\log{\eta\epsilon}.
\end{equation}

which very well fits the numerical data and at the same time is close to the approximate expression for the key rate in the case of no preparation noise (\ref{approxseries}). We also performed a comparison between a numerically calculated maximized key rate upon optimal purification, high source variance $V=10^5$, preparation noise $\Delta V \in (0,5)$ and the analytically obtained values of the key rate upon the same source variance, no purification and no preparation noise $\Delta V=0$ in the region of loss $\eta \in (0.01,0.1)$ and channel noise $\epsilon \in (0.01,0.1)$. The standard (root-mean-square) deviation of the maximized key from the analytical key rate in case of pure coherent states is approximately $6\times 10^{-5}$, which is calculated as $s=\sqrt{\frac{1}{n-1}\sum_n{\big(I_c\big|_{V =10^5}^{\Delta V=0}(\eta,\epsilon)-I_c\big|_{V = 10^5}^{T=T_{opt}}(\eta,\Delta V,\epsilon)\big)^2}}$, where $n=10^3$ is the number of points $(\eta,\Delta V,\epsilon)$ taken within the regions of parameters \cite{korn}. The average relative deviation, being the ratio of the standard deviation to the average value of the key rate for pure states $\bar{I_c}\big|_{V =10^5}^{\Delta V=0}(\eta,\epsilon)\approx 0.026$, is around  $2\%$. This divergence can be made even smaller upon further increase of source variance; for example, at $V=10^6$ it already becomes less than $1\%$.

This way we claim that by combination of optimal purification and extremely large modulation variance the key rate for collective attacks can reach the values as for the case of no preparation noise. Thus, with the optimal purification upon high modulation variance we can completely eliminate preparation noise for both types of attacks, which in the absence of purification would otherwise be destructive for the key transmission.

\section{Realistic reconciliation}
\label{recon}
In previous sections we calculated the lower bound on the secure key rate assuming that the classical postprocessing algorithms, which are aimed at deriving the secure key from the raw key, obtained by Alice and Bob, are absolutely efficient and do not reduce the key length. In practice, though the realistic error correction, which is also referred to as key reconciliation \cite{Lutk1}, has limited efficiency and is being done by the cost of the key length, thus decreasing the overall secure key rate. The negative effect of the realistic key reconciliation was shown to essentially limit the applicability of the secure key distribution with binary modulation of coherent states, especially in case of the reverse reconciliation \cite{Lutk2}. The similar effect is observed in the case of continuous-variables coding; thus, we must investigate whether the state purification is still efficient against the preparation noise if the error correction has limited efficiency.

In practical CV QKD the influence of the realistic key reconciliation is quantitatively characterized by the efficiency $\beta$, which is reducing the mutual information, available to Alice and Bob and stands for the fraction of the data, discarded in the process of error correction. Thus, in the case of collective attacks the efficient lower bound on the secure key is given by \cite{Lod2}:

\begin{equation}\label{geneff}
I_{eff}=\beta I_{AB}-\chi_{BE}
\end{equation}

Reconciliation effectiveness $\beta$ is a function of the signal-to-noise ratio (SNR), but it also depends on the algorithm, being used for the reconciliation and on the computational power, which is available for the trusted parties. Generally, the lower the SNR, the stronger the negative impact of the inefficient error correction, although for a given $\beta$, there is always some optimal value of SNR (i.e., of the modulation, applied to the coherent signal \cite{Lod2} if other parameters are fixed). So, as reconciliation efficiency $\beta$ depends on the particular technical parameters, we do not use the explicit function describing the dependence of $\beta$ on SNR in our theoretical analysis. Instead we assume that the dependence of $\beta$ on SNR is unknown, so we must keep both as the independent parameters and estimate how tolerable is the CV QKD to the preparation noise upon realistic reconciliation when we perform state purification by attenuating the signal or when the purification is not applied. As a result, we describe the area of values of SNR and $\beta$, where the method gives positive effect. Then in any particular case, when the dependence of $\beta$ on SNR is known, it can be easily found, whether given reconciliation efficiency allows improvement by the proposed purification method.

Let us consider the purely lossy channel, as the effects of both preparation noise and inefficient reconciliation are significant already in this case. In the previous sections the lower bound on the key rate $I$ was the function $I(V,\Delta V,\eta,T)$ of source variance, preparation noise, channel loss and trusted-side attenuation $T$. At the same time, SNR $\Sigma$, being the ratio of the signal variance to the overall excess noise variance at the channel output is also the function $\Sigma(V,\Delta V,\eta,T)$ of the same parameters:

\begin{equation}
\Sigma=\frac{T\eta(V-1)}{1+T\eta\Delta V}
\end{equation}

\begin{figure}
\centerline{\psfig{width=8.0cm,angle=0,file=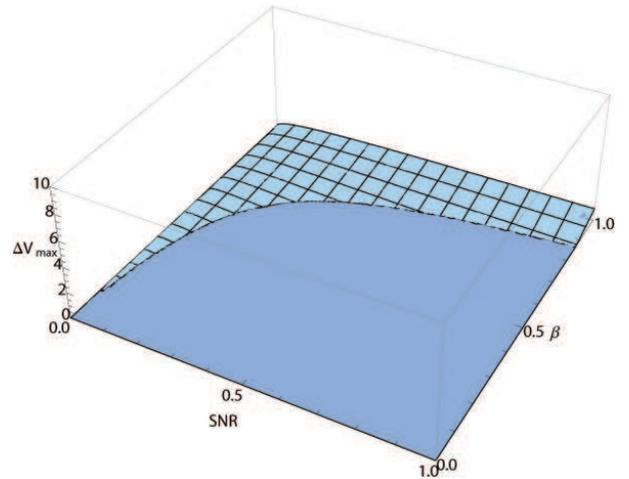}}
\caption{Threshold on the preparation noise versus reconciliation efficiency $\beta$ and SNR for any source variance $V$ and typical channel loss $\eta=0.1$ when no trusted-side purification is applied ($T=1$)}
\label{3D-dv_max_no_t}
\end{figure}

So, in order to describe the security region in terms of the preparation noise, taking into account the inefficient reconciliation, when the purification is absent i.e. $T=1$, we perform the transition $V \to V(\Sigma,\Delta V,\eta)$, recalculate the lower bound on the key rate in new terms as $I_{eff}(\Sigma,\Delta V,\eta,\beta)$, taking into account (\ref{geneff}) and can estimate the threshold values of preparation noise $\Delta V_{max}$, which turn this lower bound to zero. The threshold is presented graphically as the three-dimensional (3D) plot versus reconciliation efficiency $\beta$ and SNR for typical channel loss $\eta=0.1$ in Fig. \ref{3D-dv_max_no_t}, note that it is independent of source variance $V$, which is given by the value of SNR.

%%%%%%%%%%%%%%%%%%%%%%%%%%%%%%%%%%%%%%%%%%%%%%%%%%%%%%%%%%%%%%%%%%%%%%%%%%%%%%%%%%%%%%%%%%%
%%%%%%%%%%%%%%%%%%%%%%%%%%%%%%%%%%%%%%%%%%%%%%%%%%%%%%%%%%%%%%%%%%%%%%%%%%%%%%%%%%%%%%%%%%%%

\begin{figure}
\centerline{\psfig{width=8.0cm,angle=0,file=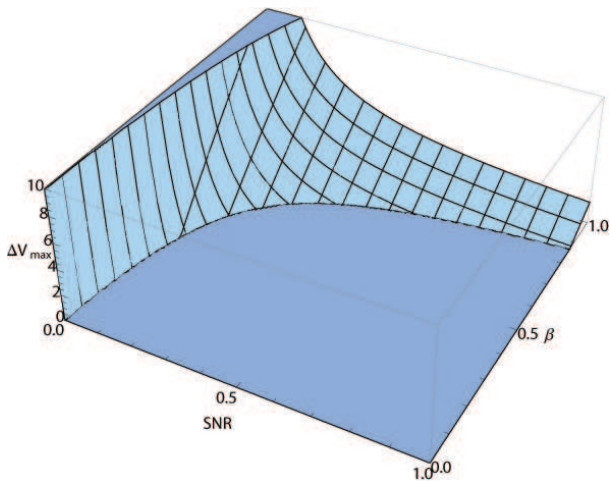}}
\caption{Threshold on the preparation noise versus reconciliation efficiency $\beta$ and SNR for realistic source variance $V=20$ and typical channel loss $\eta=0.1$ when the states are purified by attenuation $T$; values of $\Delta V_{max}$ are bounded by $10$ for the convenience of graphical representation.}
\label{3D-dv_max_t}
\end{figure}

Now let us suppose we plug in the attenuation at the trusted side, which is aimed at the purification of the signal states. Then SNR puts the constraint on attenuation $T$ as all other parameters are given and we can express $T$ as $T(\Sigma,V,\Delta V,\eta)$. Similarly, we reculaculate the lower bound on the secure key as $I_{eff}(\Sigma,V,\Delta V,\eta,\beta)$ and estimate the threshold on preparation noise $\Delta V$, when the attenuation $T$, unlike in the previous sections, is no more optimized, but is explicitly given by the SNR and other parameters. In Fig. \ref{3D-dv_max_t} we present the results as the 3D plot of the security region, describing the maximal tolerable preparation noise $\Delta V_{max}$ versus reconciliation efficiency $\beta$ and SNR, realistic source variance $V=20$ and typical channel loss $\eta=0.1$.

%%%%%%%%%%%%%%%%%%%%%%%%%%%%%%%%%%%%%%%%%%%%%%%%%%%%%%%%%%%%%%%%%%%%%%%%%%%%%%%%%%%%%%%%%%%
%%%%%%%%%%%%%%%%%%%%%%%%%%%%%%%%%%%%%%%%%%%%%%%%%%%%%%%%%%%%%%%%%%%%%%%%%%%%%%%%%%%%%%%%%%%%

Any practical scheme has its own dependence of reconciliation efficiency on SNR, which is related to the intristic parameters of the scheme, which can be graphically described by a curve in the $\beta,SNR$ plain. Thus, in order to estimate the tolerance of the particular scheme to the preparation noise and the effectiveness of state purification, one should project this curve on surfaces at Figs. \ref{3D-dv_max_no_t} and \ref{3D-dv_max_t} respectively. Nevertheless, in the general case, it is evident from both plots, that the preparation noise is significantly limiting the securirty of the scheme in the large region of SNR and reconciliation efficiency $\beta$. At the same time, the applicable purification on the trusted side can essentially improve the tolerance to the preparation noise, not reducing the applicability area in terms of SNR and $\beta$. Interestingly, this area is even increased by purification, giving a high threshold on tolerable preparation noise for low SNR and $\beta$, although the key rate in this area is very low and can be thus drastically affected by other practical imperfections.

\section{Conclusions}
We have investigated the influence of the noisy modulation on the security of the quantum key distribution scheme based on the coherent states upon realistic conditions of channel loss and detection (trusted) or channel (untrusted) excess noise. While the preparation noise was shown to be destructive to secure transmission, which reveals an essential difference between two types of trusted noise, we investigate the possibility of suppressing the preparation noise. It is shown that optimal purification on the trusted sender side drastically increases the security region in terms of the tolerable preparation noise if the channel itself is not security breaking. Thus, noisy coherent states are as useful for the secure key distribution as the pure coherent states. In the case of the high source variance, the optimal purification is shown to practically compensate the negative effect of preparation noise, which otherwise would be destructive for the transmission. The positive effect of purification is preserved also in the conditions of realistic key reconciliation with nonunity efficiency. While purification by attenuation is already being used to reduce the modulation noise and optimize the modulation for the given reconciliation efficiency, the more advanced methods out of the scope of linear operations, in particular, feedback control of the modulated signals, can be even more effective and may be the subject for further research.

%%%%%%%%%%%%%%%%%%%%%%%%%%%%%%%%%%%%%%%%%%%%%%%%%%%%%%%%%%%%%%%%%%%%%%%%%%%%%%%%%%%%%%%%%%
\medskip
\noindent {\bf Acknowledgments} The research has been supported by
Projects  No. MSM 6198959213 and No. LC06007 of the Czech Ministry
of Education and Project No. 202/07/J040 of GACR.
R.F. also acknowledges support by the Alexander von Humboldt
Foundation, V.U. thanks the Ukrainain State Fundamental Research Foundation.

%%%%%%%%%%%%%%%%%%%%%%%%%%%%%%%%%%%%%%%%%%%%%%%%%%%%%%%%%%%%%%%%%%%

\end{document}